\RequirePackage{fix-cm}

\documentclass[smallextended]{svjour3}       
\usepackage{setspace}
\usepackage{csquotes}
\usepackage{amssymb}
\usepackage{float}
\usepackage{textcomp}
\usepackage{cite}
\usepackage{amsfonts}
\usepackage[pdftex]{graphicx}
\usepackage{adjustbox}

\usepackage[cmex10]{amsmath}

\smartqed  
\usepackage{threeparttable}

  \usepackage[pdftex]{graphicx}
  \graphicspath{{pdf/}{jpeg/}}
  \DeclareGraphicsExtensions{.pdf,.jpeg,.png}
\def\makeheadbox{\relax}

\begin{document}

\title{Pre-earthquake State Identification by Micro-earthquake Spike Trains Dissimilarity Analysis
}


\author{Arash Andalib         \and
        Raheleh Baharloo      \and
         ~Jos\'{e} C. Pr\'{i}ncipe 
}


\institute{The authors are with the Electrical and Computer Engineering Department, University of Florida, Gainesville, FL 32611, USA \\
              Tel.: +1-352-392-2682\\
              Fax: +1-352-392-0044\\
              E-mails: \{andalib, principe\}@cnel.ufl.edu, baharloo@ufl.edu 
}


\maketitle

\begin{abstract}
The exact mechanisms leading to an earthquake are not fully understood and the space-time structural features are non-trivial. Previous studies suggest the seismicity of very low intensity earthquakes, known as micro-earthquakes, may contain information about the source process before major earthquakes, as they can quantify modifications to stress or strain across time that finally lead to a major earthquake. This work uses the history of seismic activity of micro-earthquakes to analyze the spatio-temporal statistical independence among the monitoring stations of a seismic network. Using point process distance measures applied to the micro-earthquakes\textquotesingle{} spike trains recorded in these stations, a pre-earthquake state is defined statistically with the aim of finding a relation between the level of dissimilarity among stations\textquotesingle{} readings and the future occurrence of larger earthquakes in the region. This paper also addresses the compatibility of this statistical approach with the Burridge-Knopoff spring-block physical model for earthquakes. Based on the results, there is evidence for an earthquake precursory state associated with an increase in spike train dissimilarity as evaluated by a statistical surrogate test.
\end{abstract}

\section{Introduction}
\label{sec:1}
Relative movements of tectonic plates lead to a slow accumulation of stress over time and along the faults near the boundaries of the plates that sometimes are exhibited by abundant micro-earthquakes. The accumulated energy is then suddenly released during an earthquake once the stress loading reaches a trigger threshold. This activity may in turn stimulate neighboring faults, developing a sequence of occurrences in space and time, to bring the dynamic medium to a new state of equilibrium \cite{bib:26}, \cite{bib:28}. Therefore, while the major energy release may seem isolated to a particular time and fault location, in general, an earthquake cannot be analyzed locally in time or space \cite{bib:1}, \cite{bib:2} and it is reasonable to look for spatial and temporal statistical dependences (aka correlation) in previous recordings, to find important information about impending earthquakes. Given the unpredictability of seismic events, the most common approach is to use a statistical approach based on point processes \cite{bib:3}, \cite{bib:27}. One difficulty is that the major earthquakes are rare and the physical processes time varying, so there is insufficient data to create accurate statistical models. One alternative is to attempt to relate micro-earthquakes with larger earthquakes, taking advantage of the larger density of micro-earthquakes and their spatial information between stations to infer statistical relationships across scales. The seismicity of micro-earthquakes in a long-term period of one decade, prior to a major earthquake, has been used to explain how the seismic or tectonic processes change ahead of large earthquakes \cite{bib:22}, but there has been no attempt in defining significant statistical criteria based on micro-earthquake activity that can be used as a precursor of large earthquakes. This work proposes to apply signal processing methods to detect abnormal seismic activities on a set of faults. In order to capitalize on the spatio-temporal structure of the micro-earthquake data a pre-earthquake state is statistically defined, also called earthquake precursory activity. The proposed method quantifies the interaction between the faults\textquotesingle{} activities obtained by seismic recordings in a distributed network across space by means of the micro-earthquakes that are produced, and uses distance measures on spike trains to evaluate their dissimilarity structure over time.

A typical seismic network includes several monitoring stations, which may be located tens to hundreds of kilometers apart from each other. These stations record local seismic activities over time. For a micro-earthquake network, the sensors are even able to detect micro-earthquakes (i.e., events with magnitudes $M\leq2$ Richter), which cannot be felt beyond several kilometers from their epicenters. While the magnitudes of these low-intensity events carry information, this work just considers the timings of micro-earthquakes so they are reduced to time-series of spike trains also called a point process. Previously, statisticians have modeled the event distribution of strong earthquakes $(M\geq6)$ over time \cite{bib:3}. Recent studies have tried to find the distribution of the number of earthquakes with magnitude larger than two on a single location to find an indicator for the temporal correlation in a single spike train \cite{bib:4}. Another approach \cite{bib:5} characterized the behavior of earthquake aftershocks $M\geq5.5$ using prototype point patterns by clustering the sequence of aftershocks of given main shocks $(7.5\leq M\leq 8)$.

This paper analyzes the dissimilarity of the spike trains from micro-earthquakes readings at the stations of a broad-band seismic network, with the aim of extracting precursors for all earthquakes $M\geq 4$ in the region, which preserves the spatial information because these small events are only detected by the closest station. Larger events, however, are recorded in almost every local station, making it unreasonable to evaluate dissimilarity between the spike trains of larger earthquakes. Here, dissimilarity is selected because it can be efficiently estimated by divergence between spike trains, while the converse (similarity) is much harder to quantify because it requires measures of statistical dependence. The Victor-Purpura (VP) distance measure \cite{bib:6}, \cite{bib:7} is selected, which is sensitive to rate and coincidence of events. Furthermore, another measure from information theoretic learning, the Cauchy-Schwarz (CS) divergence \cite{bib:8} is used, which quantifies the distance between the probability laws of two point processes in probability space. This latter metric is generally more robust than distance measures; fewer assumptions are also made when using it. The results of this study suggest that extreme dissimilarities are followed by light to significant earthquakes.

This paper is organized as follows. Sect.~\ref{sec:2} defines the pre-earthquake state and outlines how the significance of this definition is tested by surrogate method. Sect.~\ref{sec:3} describes the distance measures that are employed to analyze the spike trains. Sect.~\ref{sec:4} covers the characteristics of the seismic recordings used in this study. Sect.~\ref{sec:5} presents the results of applying dissimilarity measures to the spike trains along with how the similarities change over time. Lastly, Sect.~\ref{sec:6}, states the conclusions and offers some suggestions for the possible extensions.

\section{Methodology}\label{sec:2}
\begin{figure}[htbp]
\centering
\includegraphics[scale=0.55]{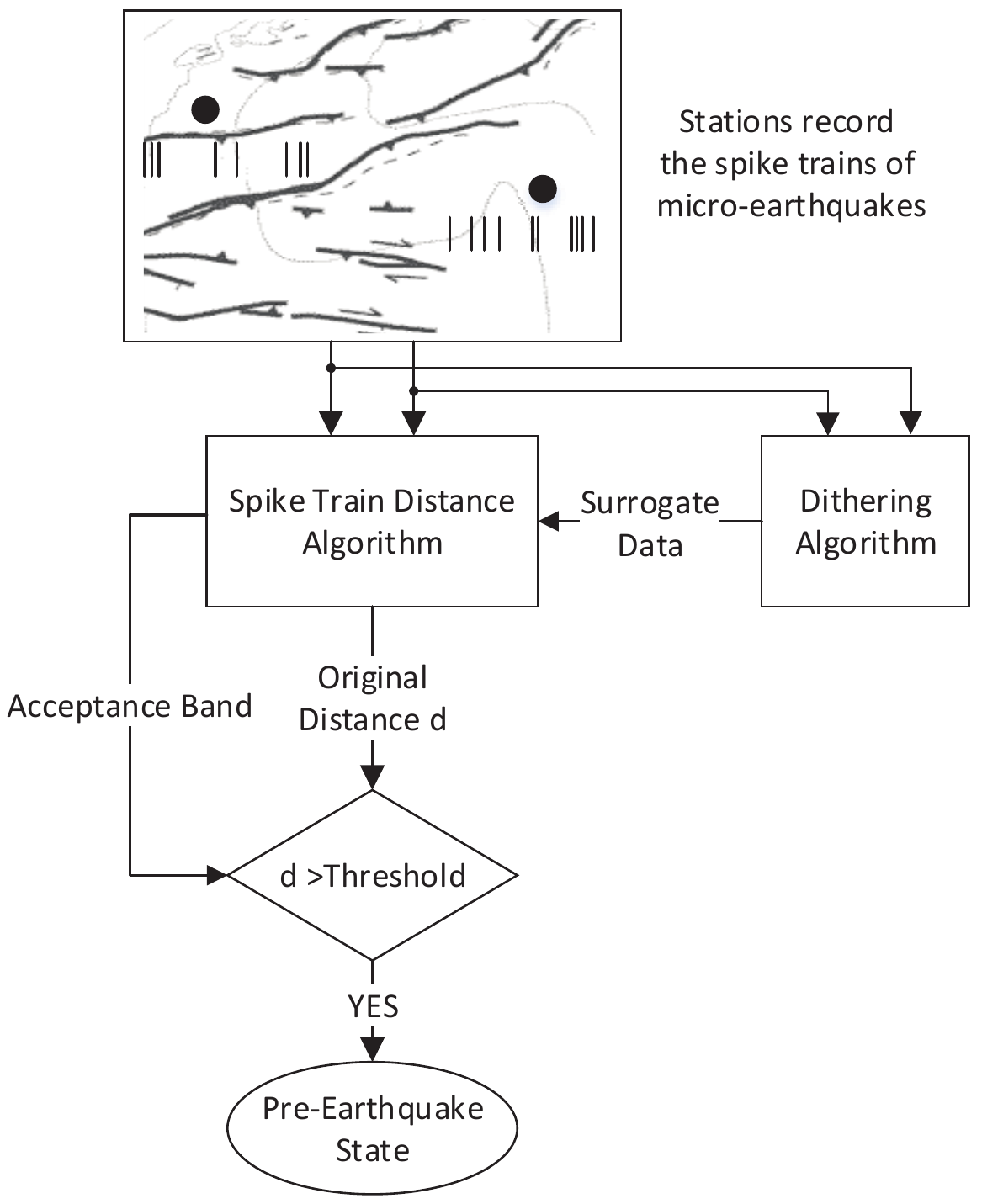}
\caption{The block diagram of the method. The upper block illustrates a typical area with some faults. Stations of a seismic network (the bullets) and corresponding spike trains are also depicted}  
\label{fig:1}
\end{figure}

Here, a statistical approach is pursued to identify abnormal behaviors ahead of large earthquakes. This approach starts with the micro-earthquakes\textquotesingle{} event times recorded in at least two stations of a sensitive broadband seismic network and applies point process distance measures to evaluate dissimilarity of the spatial seismic activity over time. Abnormal values are then identified using a statistical threshold and are labeled as the pre-earthquake state. Fig.~\ref{fig:1} illustrates the block diagram of this method, which indicates how the signal processing approaches are applied to the station measurements of the faults. This is what is discussed in detail in the rest of this paper. 
\indent

\subsection{Definition of the Pre-Earthquake State}
\label{sec:2.1}
The pre-earthquake state in this paper is defined as an occasional or durable increase in dissimilarity of spatial seismic activity, only if the amount of dissimilarity passes a statistical threshold value. The extreme (maximal) dissimilarity in the collected data quantifies a critical change in connectivity of the regional faults, which can result in a large earthquake that settles the whole system in a new state. The pre-earthquake state is based on a statistical evaluation of pairwise dissimilarity instead of physical principles. It is, however, compatible with the spring-block physical model for earthquakes and other seismic activities \cite{bib:23}. Indeed the Burridge-Knopoff spring-block model shows that the slip of one fault redefines the forces on local faults, so further slips occur and subsequently cause multiple reactionary events. When the system stress loading reaches a threshold value, a large earthquake is triggered, after which the process of relaxation begins. Based on the idea of self-organized criticality for earthquakes \cite{bib:24}, both the trigger point and relaxation point can be characterized by spatio-temporal correlation among the faults. This is exactly what the definition of pre-earthquake state is based on. 
\subsection{Statistical Test Design}
\label{sec:2.2}
To examine the hypothesis that \enquote{major earthquakes are preceded by an increase in dissimilarity of micro-earthquakes,} the null hypothesis is defined as \enquote{increases in dissimilarity are merely the results of local fluctuations and they are not related to an earthquake}. The goal is to find whether the null hypothesis can be rejected at a certain level. One option to implement the null hypothesis is to generate a Poisson point processes as a surrogate for the micro-earthquake spike trains. However, this is not easy because the rate of micro-earthquakes is continuously changing as demonstrated in Fig.~\ref{fig:3}, and this will confound the test. One widely used alternative is to synthetically create a set of spike trains by modifying the original spikes. This modification should destroy the feature of interest, which is the correlation between spike trains of earthquakes, while keeping the other statistical properties such as density intact. The surrogate data is then used to estimate the acceptance interval for normal fluctuations in dissimilarity. A dissimilarity beyond the acceptance interval is considered as an anomaly.

The dissimilarity metrics that are used in this paper do not distinguish between coincidence and changes in firing rate. Therefore, it seems quite reasonable to use surrogates that experience destruction in both coincidence and firing rates. However, if the method completely destroys the firing rate profile and reduces the inhomogeneous point process to a homogeneous one, a little change in the rate profile will lead to a false positive. The method employed to generate the surrogate data is uniform spike time dithering \cite{bib:20}, which randomly displaces spikes in a dithering window. To enforce causality in prediction, the method is slightly modified here such that dithering window always follows the original position of each spike. 

Using spike time dithering to generate surrogate data, there is a hyper parameter, which is the size of the dithering window. The dithering window needs to be long enough to destroy any coincidences in the original signal. The process of dithering also somewhat smooths the local rate profile, an effect that increases with an increase in window length. As a rule of thumb, a window length two to four times that of the window to compute the distance is recommended \cite{bib:20} for enough correlation destruction. Here, a statistical approach is used to find the optimal length of dithering window, which indeed is the minimum length that guarantees surrogates with destructed correlation. This length is equal to the lag where the original spike trains are uncorrelated. This lag corresponds to the first local minimum of cross-correlation histogram \cite{bib:30} of original spike trains. In Sect.~\ref{sec:5}, binned cross-correlation \cite{bib:29} between micro-earthquake spike trains is computed for this purpose. 

Using the dithering method explained above, two sets of surrogates are obtained from micro-earthquake data of the two stations. To produce the acceptance band, distances are computed between one-to-one pairs of surrogates from the two stations, and then are made to have identical statistical distribution via quantile normalization.

Other free parameters of each method are selected and fixed for the whole time period using 5-fold cross-validation, where each fold roughly equals one month. The positive predictive value, that is, the proportion of true positives in the positive calls, is used as the measure of fit.

\section{Spike Trains Distance Measures}\label{sec:3}
The concept of distance is conversely and strongly related to statistical dependence, which extends the concept of correlation between time series. However, unlike conventional amplitude based signals, spike train spaces are devoid of an obvious algebra. To tackle this difficulty, time binning may be used to map a spike train to Euclidean space, which allows the use of the Euclidean inner product. This process, however, has disadvantages. While binning with a coarse bin size sacrifices time precision, smaller bin sizes may keep temporal structure but are sensitive to temporal fluctuations and also suffer from dimensionality problem.

Binless measures of spike train dissimilarity have been proposed to overcome these difficulties. Most of these measures consider the spike trains to be points in an abstract metric space, proposed by Victor and Purpura \cite{bib:6}, \cite{bib:7}. The widely used time dependent approaches include the Victor-Purpura\textquotesingle s distance \cite{bib:7}, the van Rossum distance \cite{bib:9}, and the similarity measure proposed by Schreiber et al. \cite{bib:10} (see \cite{bib:11}, \cite{bib:12} for a comparison). All of these measures are dependent upon a smoothing parameter that controls the method\textquotesingle{s} sensitivity to dissimilarities in spike rate or spike coincidences. Hence, they still include a free parameter, which indicates the time precision for distance analysis, almost similar to bin size, but without time quantization.

The Victor-Purpura\textquotesingle s distance is one of the measures that is used in this work. In addition, the Cauchy-Schwarz dissimilarity measure is used, which corresponds to the correlation measure used by Schreiber et al.

\subsection{Victor-Purpura\textquotesingle s Distance}
\label{sec:3.1}
The VP distance defines the dissimilarity between two spike trains in terms of the minimum cost of transforming one spike train into the other by just three elementary operations: spike insertion, spike deletion (each with a cost of one), and shifting one spike in time to synchronize with the other. The cost of a time shift for a spike at $t_m$ to $ t_n$ is $\mathrm{q}|t_m-t_n |$, where $\mathrm{q}$ defines the time-scale with inverse time unit. The VP distance between spike trains $s_i$ and $s_j$ is defined as
\begin{equation}
d_{\mathrm{VP}}(s_i,s_j)\triangleq \min_{C(s_i \leftrightarrow s_j)} \sum_{l}{K_\mathrm{q}(t_{c_i[l]}^i,t_{c_j[l]}^j)}, \label{eqn:1}
\end{equation}
where $C(s_i\leftrightarrow s_j) $ is the set of all possible sequences of elementary operations that transform $s_i$ to $s_j$, or vice-versa, and $c_{(\cdot)}[\cdot] \in C(s_i\leftrightarrow s_j)$. That is, $c_i[l]$ denotes the index of the spike time of $s_i$ manipulated in the $l$-th step of a sequence. $K_\mathrm{q}(t_{c_i[l]}^i,t_{c_j[l]}^j)$ is the cost associated with the step of mapping the $c_i[l]$-th spike of $s_i$ at $t_{c_i[l]}^i$ to $t_{c_j[l]}^j$, corresponding to $c_j[l]$-th spike of $s_j$, or vice-versa.

Given two spike trains, each with a single spike, the distance is
\begin{equation}
K_\mathrm{q}(t_m^i-t_n^j)=\min({\mathrm{q}|t_m^i-t_n^j|,2}),\label{eqn:2}
\end{equation}
This means that the VP algorithm shifts a spike at most, $2/{\mathrm{q}}$ far from the other. Otherwise, it is cheaper to delete one of the spikes and insert another for a cost of 2. The distance $K_\mathrm{q}$ may be considered as a scaled and inverted triangular kernel applied to the spike trains \cite{bib:11}. This interpretation encourages the use of alternate  dissimilarity measures based on different kernels.
\subsection{Cauchy-Schwarz Dissimilarity}
\label{sec:3.2}
An alternative dissimilarity measure based on the Cauchy-Schwarz (CS) divergence \cite{bib:8} uses the Laplacian kernel. The kernel size $\tau$ tunes the time scale of the measure, and plays the reciprocal role of the free parameter of the VP distance $\mathrm{q}$ \cite{bib:11}. Here, by choosing a large $\tau$ the measure is more sensitive to dissimilarity in the firing rates of the spike trains, similar to the VP distance with a small $\mathrm{q}$ value. It also can be defined from the inner product of intensity functions (firing rates) of the spike trains in $L_2$.

For a spike train $s_i$ with $N_i$ spikes on the time interval $[0,T]$ and the spike times $\{t_m^i, m=1,\cdots,N_i\}$, $s_i$ can be represented as a sum of time-shifted impulses 
\begin{equation}
s_i(t)=\sum_{m=1}^{N_i}{\delta(t-t_m^i)}.\label{eqn:3}
\end{equation}

The firing rate $\lambda_{s_i}(t)$ can be estimated using a kernel smoothing representation of the spike train as
\begin{equation}
{\hat{\lambda}}_{s_i}(t)= \sum_{m=1}^{N_i}{h(t-t_m^i)},\label{eqn:4}
\end{equation}
with $h(t)$ as the smoothing kernel. This kernel needs to be non-negative valued with a unit area constraint. The memoryless cross intensity (mCI) kernel \cite{bib:13} is defined on spike trains as
\begin{equation}
I(s_i,s_j)=\int_{-\infty}^{+\infty}{{\hat{\lambda}_{s_i}}(t){\hat{\lambda}_{s_j}}(t)}dt.\label{eqn:5}
\end{equation}
Using exponential decay for kernel smoothing, the mCI kernel can be evaluated efficiently as
\begin{equation}
I(s_i,s_j)=\frac{1}{N_i N_j}\sum_{m=1}^{N_i} \sum_{n=1}^{N_j} {\kappa(t_m^i,t_n^j)},\label{eqn:6}
\end{equation}
where $\kappa(\cdot)$ is the Laplacian kernel \cite{bib:14}. The Cauchy-Schwarz dissimilarity is then defined as
\begin{equation}
d_{\mathrm{CS}}(s_i,s_j)=-\log \frac{I^2(s_i,s_j)}{I(s_i,s_i).I(s_j,s_j)}.\label{eqn:7}
\end{equation}

Mercer's theorem \cite{bib:15} implies that for the symmetric non-negative definite function $\kappa(\cdot)$ that is square integrable, the kernel has an eigen-decomposition as 
\begin{equation}
\kappa(t_m^i,t_n^j)=\langle \mathrm{\Phi}_m^i, \mathrm{\Phi}_n^j\rangle_{H_k},\label{eqn:8}
\end{equation}
where $\mathrm{\Phi(\cdot)}$ is the nonlinear mapping from the input space to the reproducing kernel Hilbert space $H_k$ induced by the kernel function. Thus, Eq.~(\ref{eqn:7}) is equivalent to
\begin{equation}
d_{\mathrm{CS}}(s_i,s_j) = -\log \frac{(\sum_{m=1}^{N_i} \sum_{n=1}^{N_j} {\langle \mathrm{\Phi}_m^i, \mathrm{\Phi}_n^j\rangle_{H_k}})^2}{(\sum_{m=1}^{N_i} \sum_{n=1}^{N_i} {\langle \mathrm{\Phi}_m^i, \mathrm{\Phi}_n^i \rangle_{H_k}})(\sum_{m=1}^{N_j} \sum_{n=1}^{N_j} {\langle \mathrm{\Phi}_m^j, \mathrm{\Phi}_n^j \rangle_{H_k}})}.
\label{eqn:9}
\end{equation}
%

The RKHS \cite{bib:21} interpretation of the CS divergence is interesting, because it does not require explicit PDF estimation. Instead, this representation provides enough space to extend the algorithm for earthquake precursory using the properties of the functional space, as will be discussed later in this paper.

\section{Experimental Data}\label{sec:4}
\begin{figure*}[!h]
\centering
\includegraphics[scale=0.7]{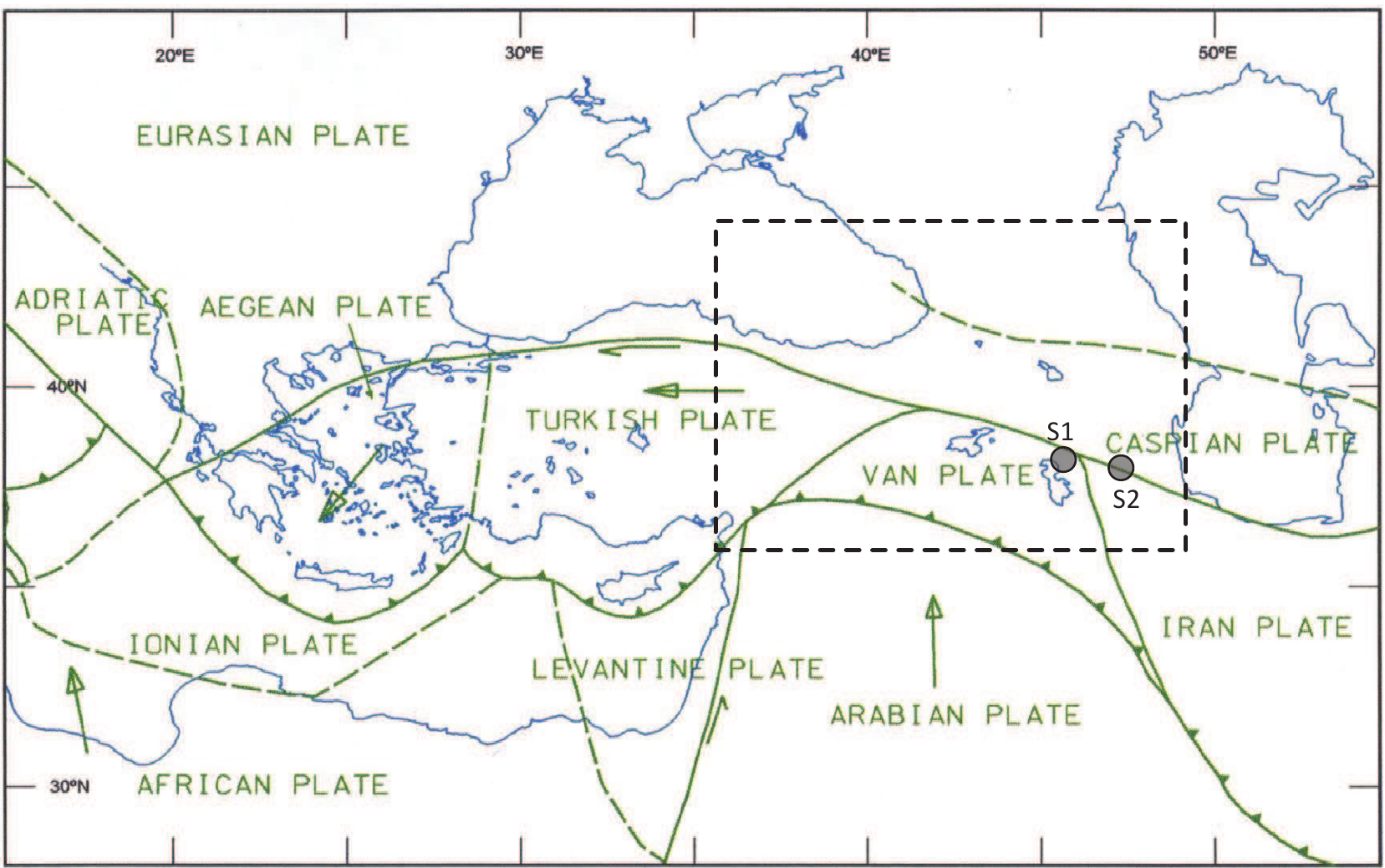}
\caption{Tectonic setting of the Alpine system \cite{bib:17}, \cite{bib:25}. The location of the stations (the bullets) and the area of interest at the intersection of Iranian, Turkish, Van, and Arabian tectonic plates (within the dashed-line rectangle) are illustrated}
\label{fig:2}
\end{figure*}

The original data is continuously recorded by surface broadband stations of the Iranian Seismological Center (IRSC). The recordings are then digitized and sent to a remote network center over satellite communication channels. At network center, the data is analyzed by a virtual seismic analyst to identify the events in real-time. The outcomes are also reviewed by an experienced seismic analyst. In this study, micro-earthquake data identified in two stations of IRSC network, Tabriz sub-network, from March 15, 2012 to August 11, 2012, are used for experiments. This is a five-month period prior to the earthquake M6.4 in northwest Iran. To define the target area, which may have precursors (warning) based on the analysis of this data set, the existing studies which explain the regional tectonic settings \cite{bib:17}, \cite{bib:25} are useful. Fig.~\ref{fig:2} illustrates the tectonic map of the Alpine system and the position of the two stations S1 and S2 (the bullets). These stations are very close to the boundaries of Iranian, Turkish, Van, and Arabian tectonic plates. The target area of this study is located at the intersection of these plates and is shown within the dashed-line rectangle (latitude:~$35.6$ to $43.1$ degrees, longitude:~$35.5$ to $49.2$ degrees). Unfortunately, not every seismic station that cover the whole area is accessible. However, this figure explains why a far earthquake in the rectangle may have a precursor provided by the two stations of this study.\\
\begin{table}[htbp]
\small
\centering
\begin{threeparttable}
\renewcommand{\arraystretch}{1.3}
\caption{The Complete Set of Earthquakes E1 to E41\tnote{a}}
\label{tab:1}       

\begin{tabular}{llllllll}
\hline\noalign{\smallskip}
   & DATE & TIME & MAG. &  LOCATION & LAT. & LON. & DEP.\tnote{b}\\
\noalign{\smallskip}\hline\noalign{\smallskip}
\textbf{E1} &\textbf{2012-03-18}	&\textbf{02:38:16}&	\textbf{4.4}& northern Iran& 36.82&	49.20& 14 \\
\textbf{E2} &\textbf{2012-03-23}&\textbf{15:43:37}&	\textbf{4.2} & eastern Turkey&	38.94&	43.62&	5 \\
E3 &2012-03-24& 06:57:47&4.2& eastern Turkey&	38.92& 43.53&	5\\
E4&	2012-03-25&	14:50:29&	4&	eastern Turkey&	39.94&	42.94&	3.4\\
\textbf{E5}&	\textbf{2012-03-26}&\textbf{10:35:32}&	\textbf{5} &	eastern Turkey&	39.17&	42.33&	5\\
\textbf{E6}&\textbf{2012-03-31}&\textbf{10:38:18}&	\textbf{4.2} &	eastern Turkey&	39.079&	43.78&	5\\
\textbf{E7}&\textbf{2012-04-04}&\textbf{09:41:40}&	\textbf{4.4} &	eastern Turkey&	38.88&	43.57&	2.6\\
\textbf{E8}&\textbf{2012-04-04}&\textbf{11:05:16}&	\textbf{4} &	Turkey-Syria border&36.93&	37.05&	8.4\\
E9&	2012-04-04&	14:18:38&	4.4 &	eastern Turkey&	39.23&	41.03&	10\\
E10&2012-04-12&	09:32:42&	4.1&	eastern Turkey&	38.69&	43.05&	2.2 \\ 
E11&	2012-04-13&	00:04:50&	4.2&Turkey-Iran border&	39.03&	44.04&	5\\
\textbf{E12}&\textbf{2012-04-13}&\textbf{04:22:08}&	\textbf{4.3}& eastern Turkey&	38.67&	43.18&	5\\
\textbf{E13}&\textbf{2012-04-18}&\textbf{23:30:58}&	\textbf{4.5} &	eastern Turkey&	38.84&	43.58&	5\\
\textbf{E14}&\textbf{2012-04-23}&\textbf{15:50:20}&	\textbf{4.1} &	Georgia&	42.32&	45.25&	10\\
\textbf{E15}&\textbf{2012-04-28}&\textbf{03:17:04}&	\textbf{4.7} &	eastern Turkey&	38.49&	40.74&	5\\
\textbf{E16}&\textbf{2012-05-07}&\textbf{04:40:27}&	\textbf{5.6} &	Azerbaijan&	41.55&	46.79&	11\\
E17&	2012-05-07&	05:38:03&	4.6&	Azerbaijan&	41.47&	46.75&	11.9\\
E18&	2012-05-07&	05:40:31&	4.6&	Azerbaijan&	41.423&	46.76&	16.6\\
E19&	2012-05-07&	08:36:24&	4.2&	Azerbaijan&	41.51&	46.77&	16.8\\
E20&	2012-05-07&	14:15:14&	5.3&	Azerbaijan&	41.55&	46.72&	11.9\\
E21&	2012-05-07&	14:36:20&	4&	Azerbaijan&	41.51&	46.73&	11.7\\
E22&	2012-05-07&	16:58:56&	4.4&	Azerbaijan&	41.56&	46.79&	14\\
\textbf{E23}&\textbf{2012-05-14}&\textbf{06:46:23}&	\textbf{4.3} &	Azerbaijan&	38.70&	48.76&	21.4\\
E24&	2012-05-14&	09:58:20&	4.1&	Azerbaijan&	41.25&	47.23&	7.4\\
E25&	2012-05-14&	15:51:02&	4&	Azerbaijan&	41.19&	47.23&	10.5\\
\textbf{E26}&\textbf{2012-05-15}&\textbf{04:54:38}&	\textbf{4.2} &	Azerbaijan&	41.54&	46.73&	17\\
E27&	2012-05-18&	14:46:35&	4.9&	Azerbaijan&	41.58&	46.76&	14.8\\
\textbf{E28}&\textbf{2012-05-18}&\textbf{14:47:22}&	\textbf{5.1} &	Azerbaijan&	41.44&	46.79&	18.1\\
\textbf{E29}&\textbf{2012-05-25}&\textbf{11:22:38}&	\textbf{4.4} &	eastern Turkey&	38.12&	38.60&	5\\
\textbf{E30}&\textbf{2012-06-05}&\textbf{16:29:48}&	\textbf{4.2}&	Azerbaijan&	41.49&	46.79&	38.3\\
\textbf{E31}&\textbf{2012-06-14}&\textbf{05:52:53}&	\textbf{5.3} &	Turkey-Syria-Iraq&37.29&42.33&5.4\\
E32&	2012-06-14&	19:17:43&	4.2&	eastern Turkey&	38.06&	42.55&	2.5 \\
\textbf{E33}&\textbf{2012-06-24}&\textbf{20:07:20}&	\textbf{4.9} &	eastern Turkey&	38.71&	43.65&	5\\
\textbf{E34}&\textbf{2012-06-25}&\textbf{20:05:59}&	\textbf{4.3} &	Azerbaijan&	41.26&	47.11&	10\\
E35&	2012-06-28&	08:39:16&	4.2&	eastern Turkey&	38.72&	43.35&	5.3 \\
\textbf{E36}&\textbf{2012-07-20}&\textbf{13:51:12}&	\textbf{4.3} &	Georgia&	42.53&	44.14&	10\\
\textbf{E37}&\textbf{2012-07-22}&\textbf{09:26:02}&	\textbf{5} &	central Turkey&	37.55&	36.38&	7.6\\
\textbf{E38}&\textbf{2012-07-24}&\textbf{22:53:39}&	\textbf{4.5} &	eastern Turkey&	38.69&	43.43&	5\\
\textbf{E39}&\textbf{2012-07-31}&\textbf{23:12:11}&	\textbf{4.1} &	eastern Turkey&	38.68&	43.05&	5\\
\textbf{E40}&\textbf{2012-08-05}&\textbf{20:37:23}&	\textbf{5} &	Turkey-Syria-Iraq&37.42&42.97&17.5\\
\textbf{E41}&\textbf{2012-08-11}&\textbf{12:23:18}&	\textbf{6.4} & northwestern Iran&38.33&46.83&	11\\
\noalign{\smallskip}\hline
\end{tabular}

\begin{tablenotes}
\item[a] Reported by USGS from March 15, 2012 to August 11, 2012, in a rectangular area (Latitude:~$35.6$ to $43.1$ degrees, Longitude:~$35.5$ to $49.2$ degrees). The main shocks, which includes 25 earthquakes, are bold highlighted
\item[b] Depths are in kilometer 
\end{tablenotes}
\end{threeparttable}
\end{table}

\indent
The rectangular search of the U.S. Geological Survey (USGS) catalog over the five-month time period includes 41 earthquakes $M\geq4$, whose characteristics are presented in Table~\ref{tab:1}. The main shocks, that is, those events that are not immediately followed by a larger earthquake, are highlighted. Among these events, there are 25 main shocks, six foreshocks and 10 aftershocks. The aftershocks are ignored, but the prediction of foreshocks is still important.\\
\indent
\begin{figure*}[!h]
\centering
\includegraphics[scale=0.65]{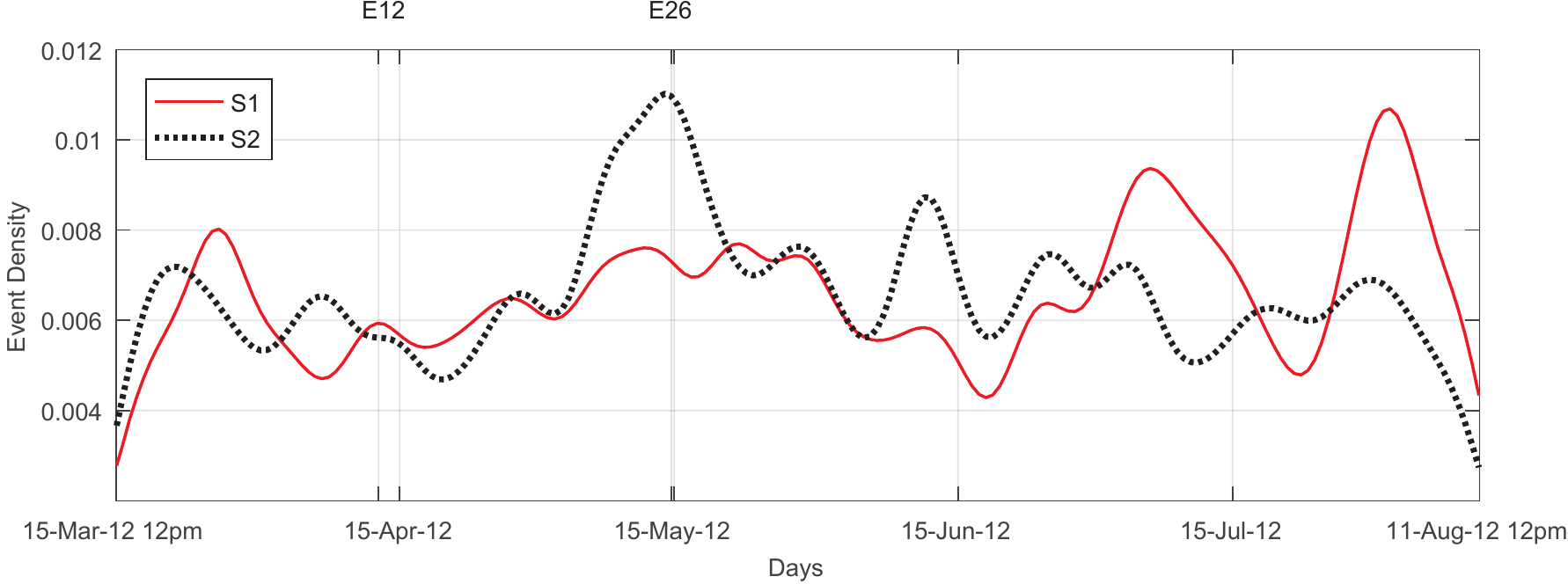}
\caption{Instantaneous rates of micro-earthquakes recorded in each station over the five-month period, created using Gaussian kernel smoothing method with optimal bandwidth of ~3 $Day$. Times of occurence of earthquakes E12 and E26 are also illustrated by vertical lines, to highlight event density differences before these two major earthquakes}
\label{fig:3}
\end{figure*}
Before computing the dissimilarity of spike trains, it is noteworthy to consider the instantaneous rate of micro-earthquakes recorded in each individual station. Here, the rate profiles are created using the Gaussian kernel smoothing method with optimal bandwidth for the kernels \cite{bib:18}. Fig.~\ref{fig:3} shows the smoothed signals. The bandwidth values are optimally set to 2.97843 $Day$ and 2.97739 $Day$ for the two stations S1 and S2, respectively. It is very important to bear in mind that the output of the smoothing filter depends on future events. Hence, it is not a causal estimator. The rate profiles may be used to understand the characteristics of the input data, however cannot provide predictability. As shown in Fig.~\ref{fig:3}, both the recorded event densities and their difference at each time instance are highly variable and demonstrate different behavior prior to major events. There are minimum and maximum differences in micro-earthquake event densities just before the major events E12 (M4.3, April 13, 2102) and E26 (M4.2, May 15, 2012), respectively. This suggests a method based on rate profile thresholding may not be enough to provide consistent information about upcoming earthquakes.

\section{Results and Discussions}\label{sec:5}
\begin{figure*}[!h]
\centering
\includegraphics[scale=0.35]{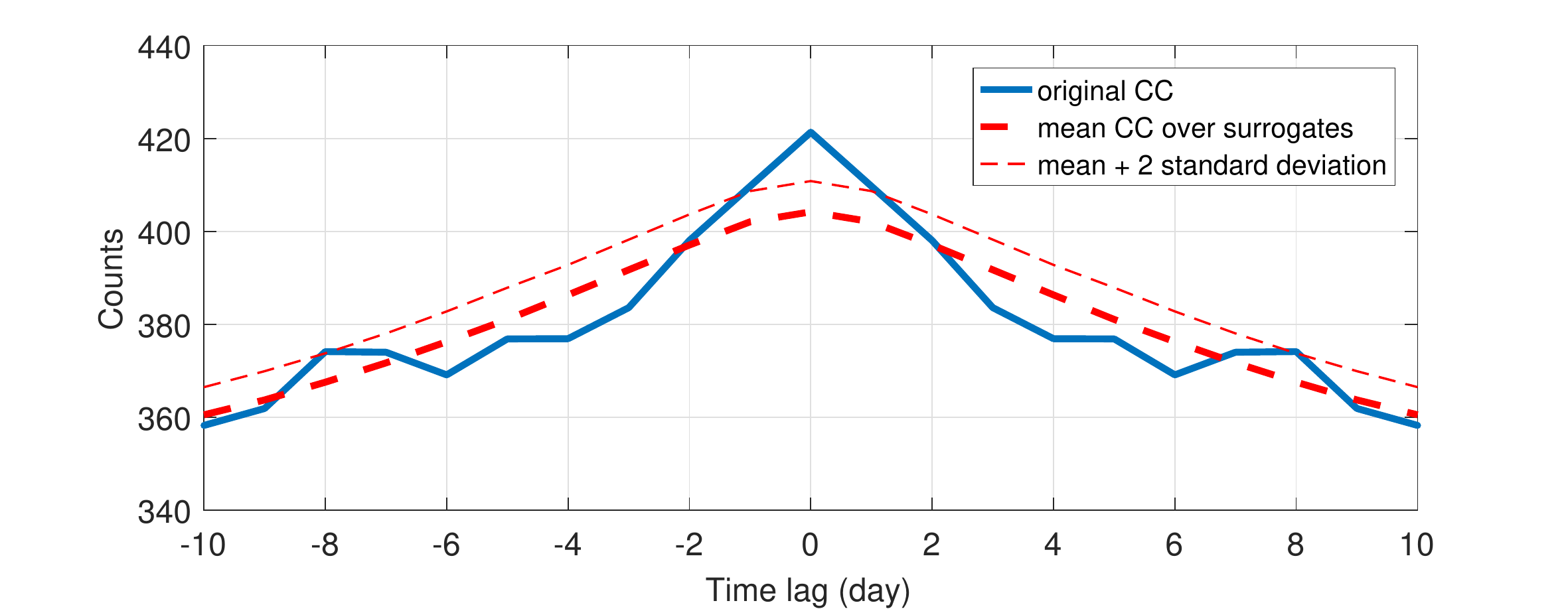}
\caption{Original and surrogate cross-correlations (CCs)}
%
\label{fig:4}
\end{figure*}

The results for applying the VP distance and CS divergence to micro-earthquakes recorded in stations S1 and S2 of Tabriz sub-network are presented. Each measure is applied to spike time vectors obtained over a sliding window. These vectors span the entire time period of interest. Hence, the output is a time resolved profile of dissimilarity of seismic activities between the two stations. Together with the distances, the one-tailed (positive) acceptance intervals from the surrogate data test at 90\%  confidence will be depicted to clearly show moments when the dissimilarities between spike trains exceed the limit of normal fluctuations on spike train structure dictated by the statistical test. The acceptance band is produced using two set of surrogates from original spike trains of stations S1 and S2, each set including 1,000 surrogates. Spike time dithering method explained in Sect.~\ref{sec:2} is used for surrogates. Following the statistical test design explained earlier, the cross-correlation (CC) of 2-day binned spike trains of micro-earthquakes is computed to find the optimal dithering window. Fig.~\ref{fig:4} depicts the original cross-correlation (the thick solid line). As expected, the cross-correlation decreases as lag increases, and has the first local minimum at 6 days. Therefore, the optimal length of dithering window is selected as 6 days. Fig.~\ref{fig:4} also depicts the mean CC over the surrogates (the thick dashed-line) and mean CC plus twice the standard deviation (the fine dashed-line). Obviously, the surrogate CC follows the shape of original CC.  

\subsection{Results for Victor-Purpura\textquotesingle s distance}\label{sec:5.1}
\begin{figure*}[!h]
\centering
\includegraphics[scale=0.67]{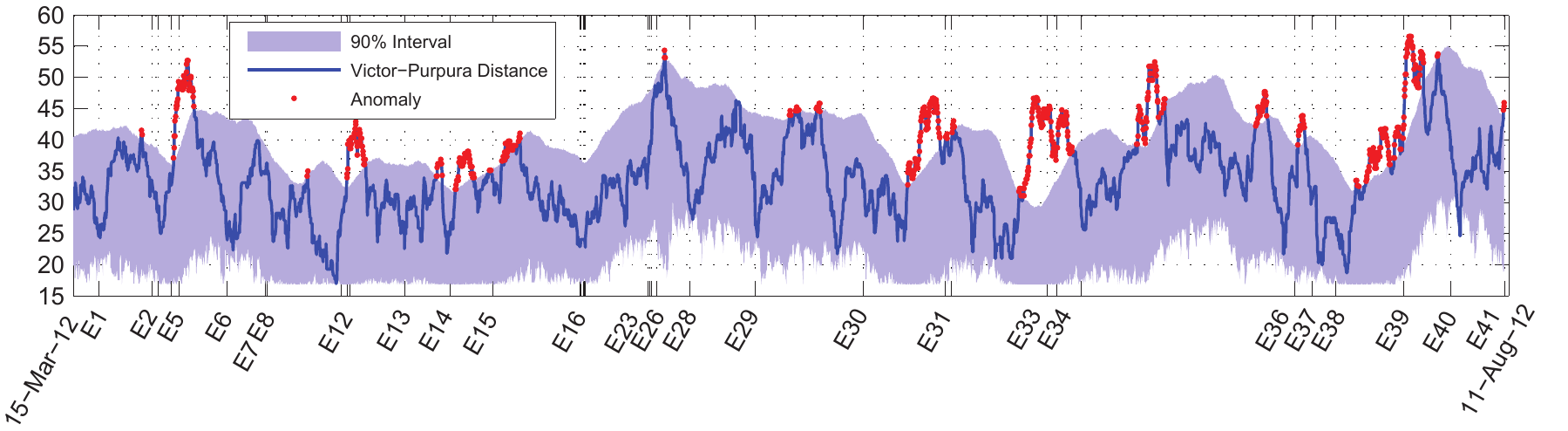}
\caption{Dissimilarity of the two stations over the five-month period, computed using the VP distance. Times of occurrence of major earthquakes $E_{1}$ to $E_{41}$ are illustrated using vertical dashed lines. Only the main shocks are labeled. The Anomalies in red are extreme dissimilarities that pass the acceptance interval}
%
\label{fig:5}
\end{figure*}

The VP distance for the original spike trains is illustrated in Fig.~\ref{fig:5}. Here, q is set to 100 $Day^{-1}$, meaning that the VP algorithm only shifts a spike which is at most 2/100 $Day$ (28.8 minutes) far from the other. The length of sliding window is set to 2 days and it slides one hour at each step. To define the one-tailed 90\%  acceptance interval, the VP distances at each step are sorted $\mathrm{VP}_{1}(l)\leq \cdots \leq \mathrm{VP}_{M}(l)$, where $l$ denotes the sliding window position, and $M$ is the number of surrogates. The lower and upper limits are then $a(l)=\mathrm{VP}_{1}(l)$ and $b(l)=\mathrm{VP}_{0.9M}(l)$, respectively.\\
\indent
\begin{figure*}[!h]
\centering
\includegraphics[scale=0.65]{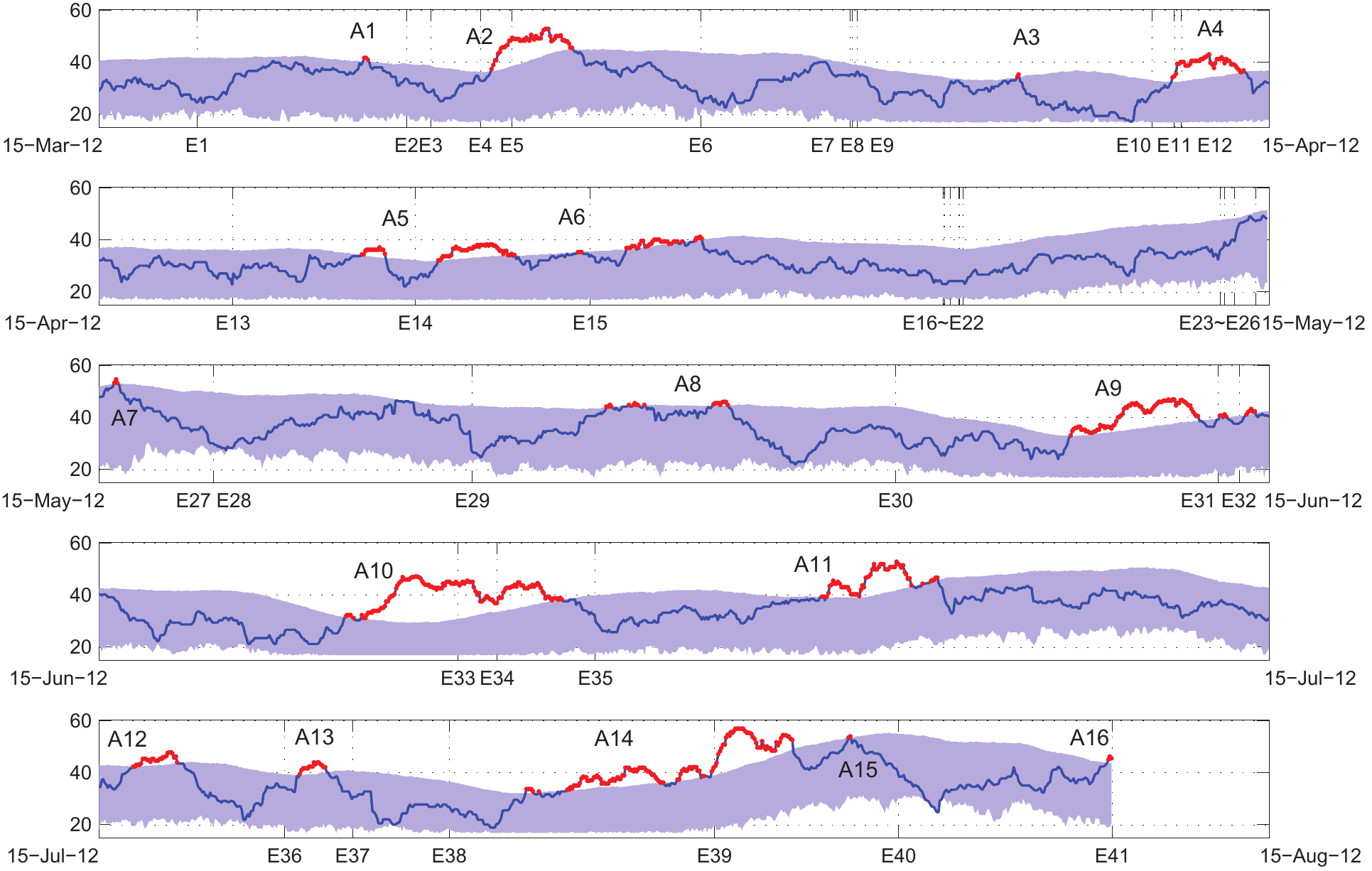}
\caption{Monthly dissimilarity of the two stations over the five-month period, computed using the VP distance. Times of occurrence of major earthquakes $E_{1}$ to $E_{41}$ are illustrated using vertical dashed lines. Anomalies are also labeled}
%
\label{fig:6}
\end{figure*}
%

\begin{table}[htbp]
\centering
\begin{threeparttable}
\renewcommand{\arraystretch}{1.3}
\caption{Confusion Matrix for Victor-Purpura Distance}
\label{tab:2}       
\small
\begin{tabular}{lll}
\hline\noalign{\smallskip}
  & True & False \\
\noalign{\smallskip}\hline\noalign{\smallskip}
Positive& 13 (E2,E12,E14,E15,E28,E31,E33,&2 (A8,A11) \\
& E34,E36,E37,E39,E40,E41)&  \\
Negative& n/a&12 (E1,E5,E6,E7,E8,E13,E16, \\
 & &E23,E26,E29,E30,E38) \\
\noalign{\smallskip}\hline
\end{tabular}
\begin{tablenotes}
\item[a] Aftershocks are ignored in this table
\end{tablenotes}
\end{threeparttable}
\end{table}
The monthly plot of Fig.~\ref{fig:6} has better visualization and the anomalies are also labeled. Comparing the anomalies with times of occurrence of major earthquakes reveals that there are extreme dissimilarities prior to 13 earthquakes, namely E2, E12, E14, E15, E28, E31, E33, E34, E36, E37, E39, E40, and E41. These anomalies may be considered as true warnings for corresponding earthquakes including E41, which is the deadly Varzaghan-Ahar earthquake M6.4 in northwest Iran. However, there are also two false positive warnings, and 12 earthquakes happen without any anomaly. The performance of the algorithm in providing efficient warnings prior to the major earthquakes is summarized in Table~\ref{tab:2}. When producing the confusion matrix in this table, the aftershocks are ignored again. Furthermore, a main shock that has a foreshock is reported as true positive, provided that the algorithm successfully predict the first earthquake, that is, the foreshock. This is the case for E12 and E28. On the other hand, while there is an anomaly before the main shock E5 it is reported false negative because the anomaly appears only after the corresponding foreshock E4.\\
\indent
\begin{table}[htbp]
\small
\centering
\begin{threeparttable}
\renewcommand{\arraystretch}{1.3}
\caption{Earthquake Precursory Performance Using the VP Distance}
\label{tab:3}       
\begin{tabular}{llll}
\hline\noalign{\smallskip}
  & & Precursory & \\ [-1ex]
\raisebox{1.5ex}{Eearthquake} &  \raisebox{1.5ex}{Anomaly} & {Time}& \raisebox{1.5ex}{Duration} \\
\noalign{\smallskip}\hline\noalign{\smallskip}
\textbf{E1}&Not Detected &No Precursory &- \\
\textbf{E2}, E3 &A1 &27 hours& 3 hours\\
E4,\textbf{E5} & Not Detected&No Precursory&-\\
\textbf{E6}& Not Detected& No Precursory& -\\
\textbf{E7}, E9& Not Detected& No Precursory& -\\
\textbf{E8}& Not Detected & No Precursory&-\\
E10, E11, \textbf{E12}& A3&86 hours&3 hour\\
\textbf{E13}&Not Detected&No Precursory& -\\
\textbf{E14}&A5& 35 hours&16 hours\\
\textbf{E15}& A6& 9 hours & 3 hours\\
\textbf{E16} to E22& Not Detected& No Precursory& -\\
\textbf{E23}& Not Detected& No Precursory& -\\
E24, E25, \textbf{E26}& Not Detected&No Precursory& -\\
E27, \textbf{E28}& A7& 65 hours& 4 hours\\
\textbf{E29}& Not Detected& No Precursory& -\\
\textbf{E30}&Not Detected& No Precursory & - \\
\textbf{E31}, E32& A9& 95 hours& 84 hours\\
\textbf{E33}, E35& A10&71 hours& Up to the event time\\
\textbf{E34}& A10& 94 hours& Up to the event time\\
\textbf{E36}& A12& 99 hours& 31 hours\\
\textbf{E37}& A13 &37 hours &20 hours\\
\textbf{E38}& Not Detected & No Precursory& -\\
\textbf{E39} &A14& 122 hours& 97\\
\textbf{E40}& A15& 34& 3\\
\textbf{E41}& A16& 3 hours& Up to the event time\\
\noalign{\smallskip}\hline
\end{tabular}

\end{threeparttable}
\end{table}

Table~\ref{tab:3} presents the precursory behavior for each earthquake. Each row of the table includes one main shock in bold, preceded (succeeded) with corresponding foreshocks (aftershocks), if any. The true positive anomalies are also reported for each group of earthquakes. Precursory time, with a mean value of 59.77 hours $(\pm 38.01)$, is the earliest warning before the event. Duration, with a mean value of 33.15 hours $(\pm 38.43)$, indicates how long the warning have been in effect in average, that is, how long the dissimilarity have been above the acceptance interval just prior to the event.
\subsection{Results for Cauchy-Schwarz divergence}\label{sec:5.2}
\begin{figure*}[!h]
\centering
\includegraphics[scale=0.67]{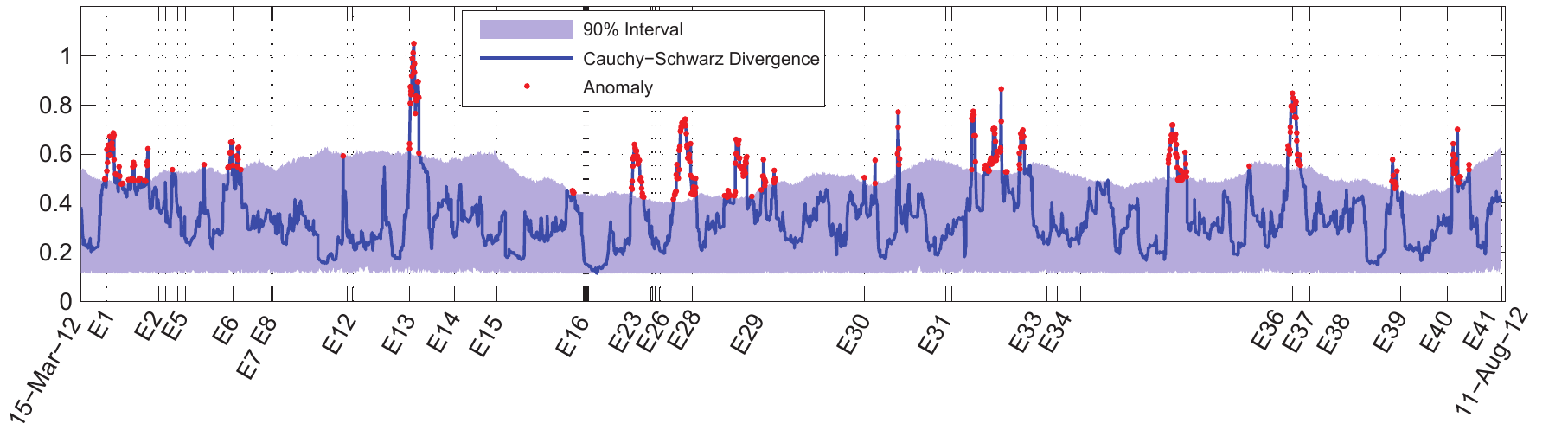}
\caption{Dissimilarity of the two stations over the five-month period, computed using the CS divergence  }
\label{fig:7}
\end{figure*}

The same experiment is repeated, using the Cauchy-Schwarz divergence instead of VP distance. The results are illustrated in Fig.~\ref{fig:7}. The kernel width is set to $\tau=2.5$ \emph{Hour}  using the cross-validation method explained earlier. Other parameters are the same as those of the VP distance.
\indent

\begin{figure*}[!h]
\centering
\includegraphics[scale=0.65]{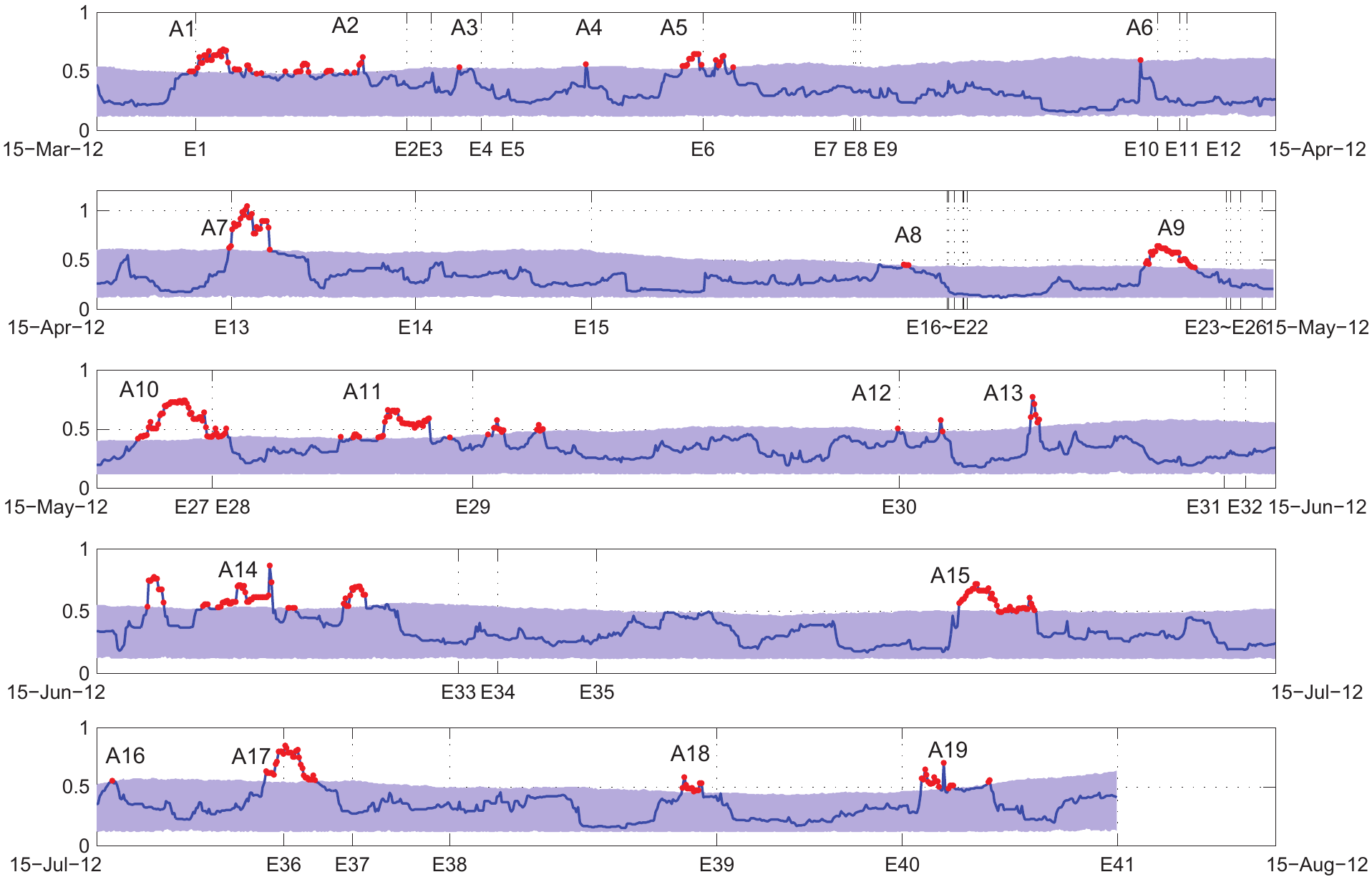}
\caption{Monthly dissimilarity of the two stations over the five-month period, computed using the CS divergence. Times of occurrence of major earthquakes $E_{1}$ to $E_{41}$ are illustrated using vertical dashed lines. Anomalies are also labeled}
\label{fig:8}
\end{figure*}

Compared with the VP distance, the CS divergence tends to be more sensitive with respect to the surrogate. The confusion matrix is presented in Table~\ref{tab:4}. The CS divergence is doing slightly better compared with the VP distance.
The monthly plot of Fig.~\ref{fig:8} indicates that there are precursors prior to 19 out of 25 main shocks and there are three false alarms (A4, A15, and A16), given an 86\% positive predictive value, which is comparable to the surrogate confidence. However, six main shocks have not any precursor.
\indent

\begin{table}[htbp]
\small
\centering
\begin{threeparttable}
\renewcommand{\arraystretch}{1.3}
\caption{Confusion Matrix for Cauchy-Schwarz Distance}
\label{tab:4}       
\begin{tabular}{lll}
\hline\noalign{\smallskip}
  & True & False \\
\noalign{\smallskip}\hline\noalign{\smallskip}
Positive&19 (E1,E2,E5,E6,E12,E13,E16,E23,E26,E28&3 (A4,A15,A16)\\
&E29,E30,E31,E33,E34,E36,E37,E39,E41)&\\
Negative&n/a &6 (E7,E8,E14,E15,E38,E40) \\
\noalign{\smallskip}\hline
\end{tabular}
\begin{tablenotes}
\item[a] Aftershocks are ignored in this table
\end{tablenotes}
\end{threeparttable}
\end{table}

Table~\ref{tab:5} discusses the precursory behavior when using the CS divergence. Here, the mean value for precursory time is 44.53 hours $(\pm 38.90)$ and the mean precursory duration is 15.71 hours $(\pm 13.63)$.\\
\begin{table}[htbp]
\small
\centering
\begin{threeparttable}
\renewcommand{\arraystretch}{1.3}
\caption{Earthquake Precursory Performance Using the CS Divergence}
\label{tab:5}       
\begin{tabular}{llll}
\hline\noalign{\smallskip}
  & & Precursory & \\ [-1ex]
\raisebox{1.5ex}{Eearthquake} &  \raisebox{1.5ex}{Anomaly} & {Time}& \raisebox{1.5ex}{Duration} \\
\noalign{\smallskip}\hline\noalign{\smallskip}
\textbf{E1}&A1 &4 hours &Up to the event time \\
\textbf{E2}, E3 &A2 &48 hours& 8 hours\\
E4,\textbf{E5} & A3&13 hours&1 hour\\
\textbf{E6}& A5& 13 hours& Up to the event time\\
\textbf{E7}, E9& Not Detected& No Precursory& -\\
\textbf{E8}& Not Detected & No Precursory&-\\
E10, E11, \textbf{E12}& A6&11 hours&1 hour\\
\textbf{E13}&A7&2 hours& Up to the event time\\
\textbf{E14}& Not Detected& No Precursory&-\\
\textbf{E15}& Not Detected& No Precursory&-\\
\textbf{E16} to E22& A8& 26 hours& 4 hours\\
\textbf{E23}& A9& 49 hours& 32 hours\\
E24, E25, \textbf{E26}& A9& 53 hours& 32 hours\\
E27, \textbf{E28}& A10& 48 hours& Up to the event time\\
\textbf{E29}& A11& 84 hours& 41 hours\\
\textbf{E30}& A12& 1 hour & 1 hour \\
\textbf{E31}, E32& A13& 121 hours& 6 hours\\
\textbf{E33}, E35& A14& 71 hours& 14 hours\\
\textbf{E34}& A14& 94 hours& 14 hours\\
\textbf{E36}& A17& 11 hours& Up to the event time\\
\textbf{E37}& A17 &54 hours &32 hours\\
\textbf{E38}& Not Detected & No Precursory& -\\
\textbf{E39} &A18& 21 hours& 13\\
\textbf{E40}& Not Detected& No Precursory& -\\
\textbf{E41}& A19& 122 hours& 21 hours\\
\noalign{\smallskip}\hline
\end{tabular}
\end{threeparttable}
\end{table}
Based on these results, it seems that the null hypothesis can be rejected using the CS measure and conclude that this method has been able to identify the pre-earthquake state based on temporal increases in dissimilarity. Although the goal in this paper is not to compare dissimilarity measures, further work is needed to explain the superior performance of the CS dissimilarity measure upon the VP distance for the surrogate test presented here. From a theoretic perspective, divergence is a stricter and a stronger statistic \cite{bib:19}, in the sense that it compares the entire probability laws. It can therefore go beyond comparing just simple statistics such as mean firing rate or spike count. It is also shown that CS dissimilarity is less sensitive to missing spikes \cite{bib:13}. Some earthquakes may not be registered by one or more monitoring stations in a seismic network, so it is important for the measure to be resistant to missing spikes and avoid deviation.
\section{Conclusion}\label{sec:6}
This paper provided a statistical representation for the pre-earthquake state by using spike train distances applied to micro-earthquakes, and tested its performance as an earthquake precursor. The spike train dissimilarity measures of Victor-Purpura distance and Cauchy-Schwarz divergence were applied to spike trains of micro-earthquakes to examine the idea that increases in dissimilarity in at least two stations may be considered as a warning for future occurrence of major earthquakes. While evidences of precursory behavior were observed using the VP distance, the CS divergence had higher performance in validating the hypothesis.

The relationship between the magnitudes of the earthquakes and the precursory behavior was not addressed in this paper. The magnitude of micro-earthquakes was ignored, and earthquakes $M\geq2$ were not incorporated in the input data set. A possible improvement is to utilize the theory of reproducing kernel Hilbert spaces (RKHS), where tensor products of multiple kernels can be defined to insert magnitude information and exploit the theory of \enquote{marked Point processes} to define distances, which takes advantage of the full information available in the earthquake catalog. Another possible extension is to insert the exact location information of the input events and relax the obligation of just labeling the input events with the stations. This may be helpful to predict the location of target earthquakes, a topic that is not addressed in this paper.

\begin{acknowledgements}
The authors of this paper are grateful to the Iranian Seismological Center for providing the data used in this study. The authors would also like to thank Dr. John F. Dewey for his helpful insight on the tectonic settings of the region of this study, and Isaac Sledge for proofreading the manuscript.
\end{acknowledgements}

\bibliographystyle{IEEEtran}
\bibliography{Earthquake_bibfile}

\begin{thebibliography}{10}
\providecommand{\url}[1]{#1}
\csname url@samestyle\endcsname
\providecommand{\newblock}{\relax}
\providecommand{\bibinfo}[2]{#2}
\providecommand{\BIBentrySTDinterwordspacing}{\spaceskip=0pt\relax}
\providecommand{\BIBentryALTinterwordstretchfactor}{4}
\providecommand{\BIBentryALTinterwordspacing}{\spaceskip=\fontdimen2\font plus
\BIBentryALTinterwordstretchfactor\fontdimen3\font minus
  \fontdimen4\font\relax}
\providecommand{\BIBforeignlanguage}[2]{{%
\expandafter\ifx\csname l@#1\endcsname\relax
\typeout{** WARNING: IEEEtran.bst: No hyphenation pattern has been}%
\typeout{** loaded for the language `#1'. Using the pattern for}%
\typeout{** the default language instead.}%
\else
\language=\csname l@#1\endcsname
\fi
#2}}
\providecommand{\BIBdecl}{\relax}
\BIBdecl

\bibitem{bib:26}
P.~Tosi, V.~D. Rubeis, V.~Loreto, and L.~Pietronero, ``Space--time correlation
  of earthquakes,'' \emph{Geophysical Journal International}, vol. 173, no.~3,
  pp. 932--941, 2008.

\bibitem{bib:28}
\BIBentryALTinterwordspacing
A.~V.~M. Herz and J.~J. Hopfield, ``Earthquake cycles and neural
  reverberations: Collective oscillations in systems with pulse-coupled
  threshold elements,'' \emph{Phys. Rev. Lett.}, vol.~75, pp. 1222--1225, Aug
  1995. [Online]. Available:
  \url{http://link.aps.org/doi/10.1103/PhysRevLett.75.1222}
\BIBentrySTDinterwordspacing

\bibitem{bib:1}
V.~P. Plagianakos and E.~Tzanaki, ``Chaotic analysis of seismic time series and
  short term forecasting using neural networks,'' in \emph{Proc. IEEE
  International Joint Conference on Neural Networks}, vol.~3, 2001, pp.
  1598--1602.

\bibitem{bib:2}
R.~J. Geller, D.~D. Jackson, Y.~Y. Kagan, and F.~Mulargia, ``Earthquakes cannot
  be predicted,'' \emph{Science}, vol. 275, pp. 1616--1617, 1997.

\bibitem{bib:3}
Y.~Ogata, ``Statistical models for earthquake occurrence and residual analysis
  for point processes,'' \emph{Journal of American Statististical Association},
  vol.~83, pp. 9--27, 1988.

\bibitem{bib:27}
Y.~Kagan and D.~Vere-Jones, ``Problems in the modelling and statistical
  analysis of earthquakes,'' in \emph{Athens Conference on Applied Probability
  and Time Series Analysis}.\hskip 1em plus 0.5em minus 0.4em\relax Springer,
  1996, pp. 398--425.

\bibitem{bib:22}
R.~Stefansson, \emph{Advances in earthquake prediction}.\hskip 1em plus 0.5em
  minus 0.4em\relax New York: Springer, 2011, pp. 92--93.

\bibitem{bib:4}
M.~Karsai, K.~Kaski, A.~L. Barabási, and J.~Kertész, ``Universal features of
  correlated bursty behavior,'' \emph{Scientific Reports 2}, 2012.

\bibitem{bib:5}
F.~P. Schoenberg and K.~E. Tranbarger, ``Description of earthquake aftershock
  sequences using prototype point processes,'' \emph{Environmetrics}, vol.~19,
  pp. 271--286, 2011.

\bibitem{bib:6}
J.~D. Victor and K.~P. Purpura, ``Nature and precision of temporal coding in
  visual cortex: a metric-space analysis,'' \emph{Journal of Neurophysiology},
  vol.~76, pp. 1310--1326, 1996.

\bibitem{bib:7}
------, ``Metric-space analysis of spike trains: theory, algorithms and
  application,'' \emph{Network 8}, pp. 127--164, 1997.

\bibitem{bib:8}
J.~C. Pr\'{i}ncipe, \emph{Information Theoretic Learning: Renyi\textquotesingle
  s Entropy and Kernel Perspectives}.\hskip 1em plus 0.5em minus 0.4em\relax
  Springer, 2010.

\bibitem{bib:23}
R.~Burridge and L.~Knopoff, ``Model and theoretical seismicity,'' \emph{Bull.
  Seismol. Soc. Am.}, vol.~57, no. 341, 1967.

\bibitem{bib:24}
Z.~Olami, H.~J.~S. Feder, and K.~Christensen, ``Self-organized criticality in a
  continuous, nonconservative cellular automaton modeling earthquakes,''
  \emph{Phys. Rev. Lett.}, vol.~68, pp. 1244--1247, 1992.

\bibitem{bib:20}
S.~Louis, C.~Borgelt, and S.~Grun, ``Generation and selection of surrogate
  methods for correlation analysis,'' in \emph{Analysis of Parallel Spike
  Trains}, S.~Grun and S.~Rotter, Eds.\hskip 1em plus 0.5em minus 0.4em\relax
  Springer Series in Computational Neuroscience, 2010.

\bibitem{bib:30}
D.~H. Perkel, G.~L. Gerstein, and G.~P. Moore, ``Neuronal spike trains and
  stochastic point processes: Ii. simultaneous spike trains,''
  \emph{Biophysical journal}, vol.~7, no.~4, pp. 419--440, 1967.

\bibitem{bib:29}
A.~R.~C. Paiva, I.~Park, and J.~C. Pr{\'i}ncipe, ``Reproducing kernel hilbert
  spaces for spike train analysis,'' in \emph{ICASSP}, 2008.

\bibitem{bib:9}
M.~C.~W. van Rossum, ``A novel spike distance,'' \emph{Neural Computation},
  vol.~13, pp. 751--763, 2001.

\bibitem{bib:10}
S.~Schreiber, J.~M. Fellous, J.~H. Whitmer, P.~H.~E. Tiesinga, and T.~J.
  Sejnowski, ``A new correlation based measure of spike timing reliability,''
  \emph{Neurocomputing}, vol.~52, pp. 925--931, 2003.

\bibitem{bib:11}
A.~R.~C. Paiva, I.~M. Park, and J.~C. Pr\'{i}ncipe, ``A comparison of binless
  spike train measures,'' \emph{Neural Computing and Applications}, vol.~19,
  pp. 405--419, 2010.

\bibitem{bib:12}
T.~Kreuz, D.~Chicharro, M.~Greschner, and R.~G. Andrzejak, ``Time-resolved and
  time-scale adaptive measures of spike train synchrony,'' \emph{Journal of
  Neuroscience Methods}, vol. 195, pp. 92--106, 2011.

\bibitem{bib:13}
I.~M. Park, ``Capturing spike train similarity structure: point process
  divergence approach,'' Ph.D. dissertation, Univ. of Florida, 2010.

\bibitem{bib:14}
A.~R.~C. Paiva, I.~Park, and J.~C. Pr\'{i}ncipe, ``A reproducing kernel hilbert
  space framework for spike trains,'' \emph{Neural Computation}, vol.~21,
  no.~2, pp. 424--449, 2009.

\bibitem{bib:15}
J.~Mercer, ``Functions of positive and negative type, and their connection with
  the theory of integral equations,'' \emph{J. Philos. Trans. Royal Society of
  London}, vol. 209, pp. 415--446, 1909.

\bibitem{bib:21}
N.~Aronszajn, ``The theory of reproducing kernels and their applications,''
  \emph{Cambridge Philo. Soc. Proc.}, vol.~39, pp. 133--153, 1943.

\bibitem{bib:17}
J.~F. Dewey, W.~C. Pitman, W.~B.~F. Ryan, and J.~Bonnin, ``Plate tectonics and
  the evolution of the alpine system,'' \emph{Geol. Soc. Amer. Bull.}, vol.~84,
  pp. 3137--3180, 1973.

\bibitem{bib:25}
M.~Erdik, Y.~A. Biro, T.~Onur, K.~Sesetyan, and G.~Birgoren, ``Assessment of
  earthquake hazard in turkey and neighboring,'' \emph{Annals of Geophysics},
  vol.~42, no.~6, pp. 1125--1138, 1999.

\bibitem{bib:18}
H.~Shimazaki and S.~Shinomoto, ``Kernel bandwidth optimization in spike rate
  estimation,'' \emph{Journal of Computational Neuroscience}, vol.~29, no. 1-2,
  pp. 171--182, 2010.

\bibitem{bib:19}
I.~M. Park, S.~Seth, M.~Rao, and J.~C. Pr\'{i}ncipe, ``Strictly positive
  definite spike train kernels for point process divergences,'' \emph{Neural
  Computation}, vol.~24, no.~8, 2012.

\end{thebibliography}

%
%

\end{document}